# Multidisciplinary Engineering Models: Methodology and Case Study in Spreadsheet Analytics

David Birch[a] *, Helen Liang[b], Paul H J Kelly[a], Glen Mullineux[b],
Tony Field[a], Joan Ko[c], Alvise Simondetti[c]
[a] Imperial College London  [b] University of Bath  [c] Arup
* David.Birch@Imperial.ac.uk

## Abstract

This paper demonstrates a methodology to help practitioners maximise the utility of complex *multidisciplinary* engineering models implemented as spreadsheets, an area presenting unique challenges. As motivation we investigate the expanding use of Integrated Resource Management (IRM) models which assess the sustainability of urban masterplan designs. IRM models reflect the inherent complexity of multidisciplinary sustainability analysis by integrating models from many disciplines. This complexity makes their use time-consuming and reduces their adoption.

We present a methodology and toolkit for analysing multidisciplinary engineering models implemented as spreadsheets to alleviate such problems and increase their adoption. For a given output a relevant slice of the model is extracted, visualised and analysed by computing model and interdisciplinary metrics. A sensitivity analysis of the extracted model supports engineers in their optimisation efforts. These methods expose, manage and reduce model complexity and risk whilst giving practitioners insight into multidisciplinary model composition. We report application of the methodology to several generations of an industrial IRM model and detail the insight generated, particularly considering model evolution.

## 1. Introduction

Many multidisciplinary engineering models are implemented as spreadsheets for ease of construction, modification and portability amongst practitioners. While many benefits are realised by an integrated spreadsheet based model, there are some inherent difficulties common to many engineering models. To demonstrate these challenges, we consider those within the urban masterplanning community. Urban masterplanning is the process of creating a coherent design for the development of a campus, suburb, city or region. It spans not only architecture but the disciplines involved in the implementation of changes to the built environment such as acoustics and water supply.

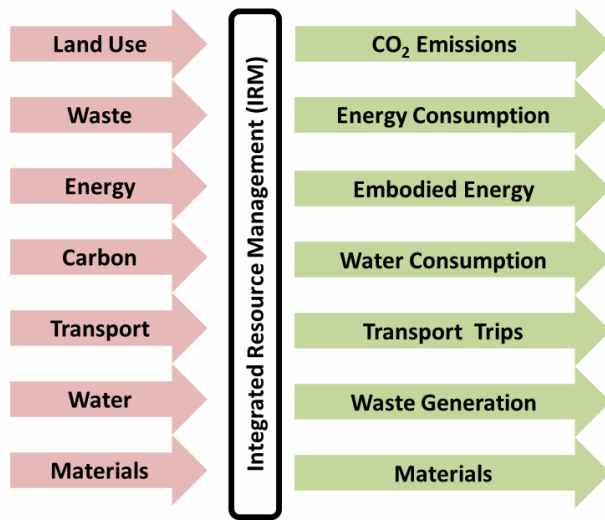

Figure 1 Conceptual model of an Integrated Resource Management (IRM) Model [Ayaz08]. Sustainability models from many disciplines are integrated to form a coherent model for assessing urban masterplans [Page08].

Increasing requirements for managing environmental impact have led to demand for interdisciplinary modeling of sustainability metrics such as annual per capita carbon emissions in order to benchmark and improve designs. These drivers have been unified by Integrated Resource Management (IRM) models [Kepran, 2002; Ayaz, 2008; Page, 2008] which integrate models from each discipline into a coherent assessment tool..The challenges encountered in such models motivate this work and are discussed in the next section. This paper presents the following contributions to address these issues.

- We present a methodology and tool suite for systematic, automatic analysis of large spreadsheet-based models with novel metrics to assess internal communication and integrated sensitivity analysis to aid practitioners in optimisation.

- We apply this methodology with a focus upon multidisciplinary engineering assessment models, a model type not widely studied within literature.

- We demonstrate the methodology's application through practical case studies with an industrial multidisciplinary sustainability model, identifying insight for practitioners and study model evolution over three model generations.

## 2. Motivation

In this paper we consider Arup's IRM model [Ayaz, 2008; Page, 2008] as an example of a complex spreadsheet based interdisciplinary engineering model. Arup is a global engineering consultancy and their IRM model is used frequently worldwide on a wide-range of projects. We now consider some of the challenges inherent to all such interdisciplinary engineering models.

As shown in Figure 2, Arup's IRM model consists of several different discipline specific sub-models including energy demand, energy supply, passenger transport and land-use. Each discipline has a data input model and an output model which calculates sustainability metrics such as annual energy demand. These input/output model pairs strongly rely, not only, upon each other, but also upon the other disciplines' input and output models. For example, the energy supply sub-model uses inputs from the land-use input sub-model and the outputs of the energy demand model.

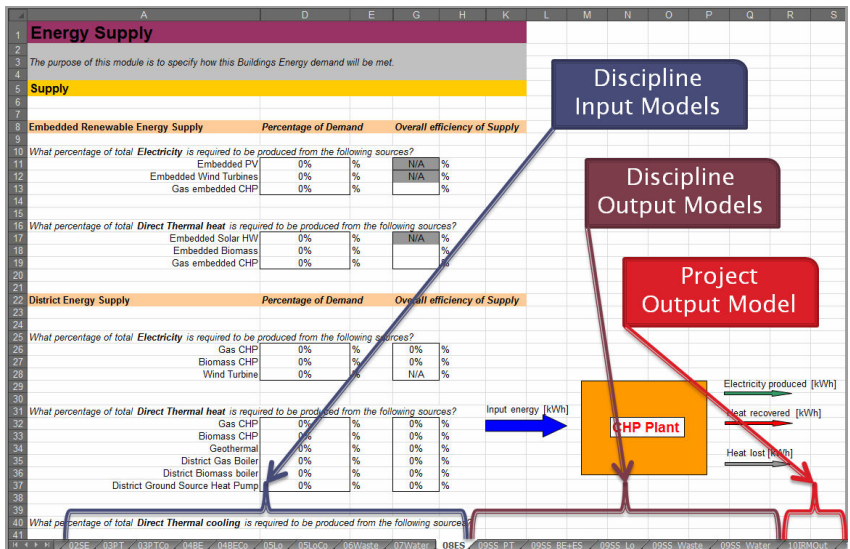

**Figure 2 Arup's IRM model [Page08], is implemented as a Microsoft Excel spreadsheet. Each discipline has an input and an output model in its own worksheet. A single project metrics dashboard is provided.**

This creates a complex interrelated web of models which reflects the physical complexity of sustainability concerns. In the centre of this web is a project-specific sustainability dashboard calculates summary metrics using information from all disciplines' input and output models. This complexity is a requirement of faithful modeling and is a common feature of many engineering analysis models. This class of models, in contrast to more traditional spreadsheet based models such as tax calculators, face particular challenges:

- **Model Complexity** - Such models are by their nature complex due to the strong coupling between already intricate discipline models which must become facsimiles of real life complexity. This leads to difficulty in gaining an accurate overview of the whole model and to

- **Data Requirements** - Due to this complexity, engineering models frequently contain large data requirements. Our analysis identified 933 (see Section 10) separate design or analysis variables required for the carbon calculation of Arup's IRM model, ranging from the total floor area of residential buildings to the $CO_2e$ emissions for disposing of electronic equipment. The time taken to gather, process and enter the required information is a major cost in applying such models.
- **Implicit Knowledge** - Such engineering knowledge is difficult to formalise, being built up as an informal set of good practice over time. Formalising and modifying this implicit knowledge is challenging, particularly when there is limited documentation (e.g. spreadsheet formulas).
- **Interdisciplinary Communication** - Within multidisciplinary models, each discipline has its own nomenclature which must be communicated to the other disciplines involved. Given the limited documentation in many spreadsheets, this may result in the same figure to be included under different names in different units.
- **Project Adaptation** - In contrast with many fixed purpose models implementing a clear specification (e.g. tax law), most engineering models, whilst trying to be as general as possible, often require some tweaking to fit the exact nature of the task at hand. Due to its scope, an IRM model often requires adaptation to each project for the following reasons:
  - **Models too broad** - A model's data requirements are large and can prove broader than the scope of the project, especially during early design stages. This leads to difficulty in fulfilling all the data requirements.
  - **Models too narrow** - A common cause of model adaptation is to meet project specific concerns. For example the inclusion of irrigation and grey water recycling is critical in water stressed areas but is rarer in more temperate climates and so may need to be added to the model.
  - **Cause and Effect unclear** - Project adaptation for these reasons is a difficult activity - the scale of the model makes identification of cause and effect between an input to be modified and the final sustainability metric difficult to determine, especially because of the interrelated nature of multidisciplinary models.
- **Difficulty of Optimisation** - Once an engineering model is applied to a project the most common use is to create a number of design improvement recommendations. This is difficult since it depends on understanding both the overview and the detail of the model. This requires high levels of implicit knowledge in varying assumptions and understanding the flow of cause and effect across multiple discipline models to identify the handful of most advantageous steps that could be taken to improve the design.
- **Implementation** - Whilst spreadsheet based models are common and support ease of use and modification (a survey undertaken by the authors identified around 1,000 engineering analysis models in use within a large engineering firm). There is a growing body of evidence that spreadsheet models in common with other large software products are likely to contain errors at unacceptable rates. A good summary of the current evidence is available in [Panko, 2008].

understanding how a single discipline's model functions; especially outside of a practitioner's area of expertise.

In summary, there are clear obstacles in the use of all spreadsheet based multidisciplinary engineering models. This paper demonstrates the value of model analysis tools to support practitioners in their information intensive tasks.

## 3. Methodology

As proposed in [Liang, 2011] with application to the design process, we propose and demonstrate an Extraction and Analysis Methodology (EAM) consisting of a series of techniques to help expose, reduce and manage model complexity. In this paper we explore the impact on multi-disciplinary engineering models. We demonstrate insight into multidisciplinary model composition and show value for designers in quickly focusing efforts into optimisation.

The methodology has the following steps:
1. **Obtain** - Model and project objectives.
2. **Define** - Key Performance Indicators (KPIs) of interest to the project.
3. **Extract** - Slice model to expose and reduce complexity to produce a smaller model computing only the KPIs of interest.
4. **Analyse - Visualise** - Visualise model to aid comprehension and show cause and effect.
5. **Analyse - Metrics** - Compute metrics on calculation model to give insight into model composition.
6. **Optimise** - Set variable ranges to formalise implicit knowledge enabling sensitivity analysis to give insight and focus optimisation effort.

The benefits of this methodology are in the value they provide to the practitioner. Firstly, by reducing the problem size and allowing visualisation to enable interactive exploration of cause and effect. Secondly, by providing metrics and insight into the multidisciplinary composition of the model we show the interaction of various disciplines. Finally, a sensitivity analysis provides further insight and focuses design effort enabling faster optimisation. The methodology also aids model development and evolution as the models are adapted to new projects. Similarly it helps mitigate the risks associated with spreadsheet based modelling particularly during the modification and optimisation stages of use.

## 4. Related Work

Studies have identified the presence [Panko, 2008; Clermont, 2005] and frequency [Blayney, 2006] of spreadsheet errors. We know that the majority of modellers do not have formal training in spreadsheet based modelling [Panko, 2008]. A body of literature has developed aiming to formalise a taxonomy of spreadsheet modelling bugs [Panko, 2010]. The risks of these errors are commonly underestimated and few users of spreadsheets consider the risks of such errors [Blayney, 2006]. Indeed very few practitioners consider that they need tools for debugging their models. There have been a number of studies into auditing tools for spreadsheets (e.g. [Blayney, 2006] for tax purposes). Historically there has been much interest in deriving visualisations based on the calculation graph of a spreadsheet [Kankuzi, 2008; Shiozawa, 1999]. Several visualisation tools have been proposed to avoid costly errors.

The novelty of our approach is that rather than treating a spreadsheet as simply a software artefact we consider the insight each step and tool in our methodology can generate for the model maintainer with a view to aiding them as they optimise a design. This is particularly a challenge for engineering models as oppose to financial models which have previously been the focus of research. These engineering models through their constant evolution and adaptation to projects present new research challenges. Particularly we propose a life-cycle methodology for the use of such tools by practitioners. We also consider for the first time, the challenges that a multidisciplinary model brings to the challenge of spreadsheet engineering. For example, considering approaches for assessing multidisciplinary communication within models (Sections 7 and 8). We also consider how sensitivity analysis may be performed in large spreadsheet based models. This is enabled through our extraction and analysis methodology and has the potential to generate substantial insight for practitioners as evidenced in Section 10. Finally, we consider the evolution of complex models as they are developed and applied to projects. As discussed in Section 11 this is a great source of insight into the model and a future research challenge.

## 5. Model Extraction

The first stage of the methodology is to extract a slice of the model from Excel. Slicing a model or computer program is a well-known technique [Weiser, 1981] that allows consideration of only the portion of the model involved. In this context, slicing extracts only spreadsheet cells involved in the calculation of particular outputs, reducing the model size and complexity.

We recursively extract cells by starting from the outputs of interest (e.g. annual per capita carbon emissions), read their formula parsing them for references to other cells, recursively extracting these until no more cells are referenced. We used a mathematical expression evaluation library NCalc and modified the grammar to be compatible with Microsoft Excel formulas and implemented a subset of Excel functions allowing internal evaluation of formulas to enable validation of analysis. In contrast with many other approaches [Reichwein, 1999; Shiozawa, 1999; Kankuzi, 2008] this formula parsing approach enables us to gain insight within formulas, for example differentiating cells referenced from arithmetic from table lookup functions which reference hundreds of cells. This enables simplification of the extracted model slice and resultant graph of cells. We also extract cell values and names to aid comprehension of visualisations, metrics and sensitivity analysis. We find model slicing a key contribution to the comprehensibility of the resulting calculation graph as is shown in the next section.

## 6. Visualisation

Taking inspiration from [Shiozawa, 1999; Kankuzi, 2008; Hermans, 2011], our methodology includes a calculation graph visualisation. We present cells and ranges as nodes in the graph and references between them as edges. We colour the nodes according to which discipline model they originate from, giving insight to discipline communication. Additionally, we support interactive exploration of the calculation graph under several layouts each highlighting different aspects of the graph.

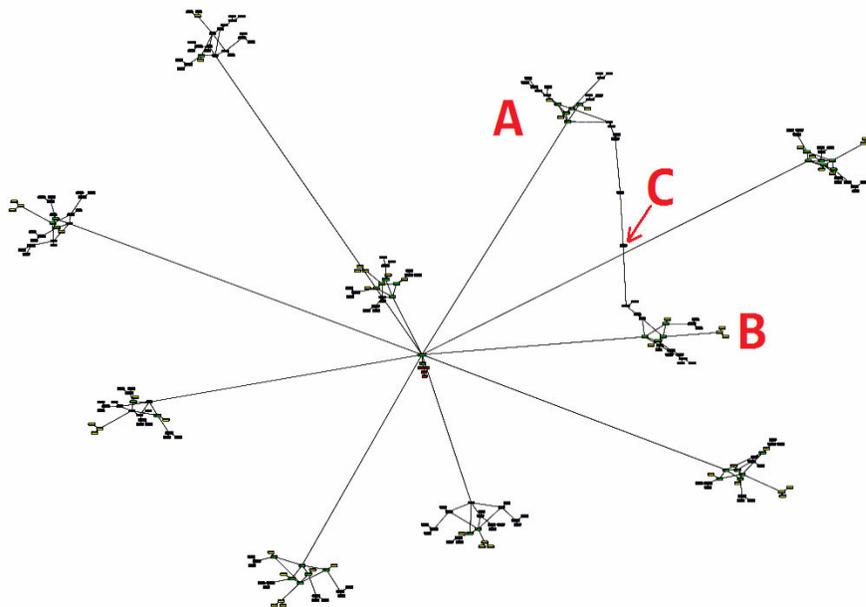

For example Figure 3 highlights the complexity of $CO_2$e emissions per capita per annum for external transport calculation within Arup's IRM model. This model slice contains 255 cells and is visualised with a linlog energy force model highlighting strongly connected sub-graphs. Hence we see ten sub-graphs (calculations) feeding into the metric (the central node in the graph), corresponding to the calculation of carbon emissions for ten modes of transport.

**Figure 3 Calculation graph for CO2e emissions percapita per annum for external transport. Layout highlights sub-calculations for ten modes of transportation.**

An interesting graph anomaly is that two sub-calculations are connected ("A" and "B"). Upon selecting node "C" linking both calculation clusters ("A" and "B"), a list of the ways this cell is used in the calculation of the metric is generated. Further investigation shows the input value ("C") to represent the $CO_2e$ emissions for diesel buses per passenger kilometre. This is used in the calculation of both the *bus* and *coach* modes of transport ("A" and "B"). This is unexpected since coaches are normally have around a quarter of the $CO_2e$ emissions of buses. This implies the carbon emissions for coaches could be overestimated in the model. This issue was reported to the IRM engineers who agreed the issue was unexpected and had been fixed in later versions of the model but could have been an assumption carried over from a previous project where coaches and buses have similar $CO_2e$ emissions on small islands.

This demonstrates the utility of slicing and visualisation tools to aid understanding and examination of complex engineering models.

## 7. Model Metrics

Having extracted a slice of a multidisciplinary engineering model various graph metrics can be automatically calculated to give insight into the multidisciplinary composition of the calculation model. In contrast with previous approaches we consider the value of worksheet level metrics rather than formula level metrics [Hodnigg, 2008; Hermans, 2012], particularly because of the relationship with the disciplines they represent.

Firstly we can partition the calculation graph by discipline and gain a measure of their complexity via the cell count and number of inputs in their partition. This is shown in Figure 4 which also shows the average valency (average number of cells each cell references and is referenced by). More references show more complexity and interconnectivity which although harder to maintain, may model reality more accurately. Arup's IRM model's carbon calculation has 2,357 nodes with average valency of 2.89. In Figure 4 we see the model's focus upon Energy and Passenger Transport with the Transport input models and the Energy output models containing most complexity and interconnectivity.

| Discipline Model | Cell Counts | Inputs | % Inputs | Average Valency |
|---|---|---|---|---|
| Land Use (LU) | 38 | 24 | 63% | 3.24 |
| Socio Economic (SE) | 38 | 23 | 61% | 1.87 |
| Passenger Trans (PT) | 210 | 180 | 86% | 1.57 |
| Pass Trans Coeff (PTCo) | 140 | 99 | 71% | 2.44 |
| Energy Demands (ED) | 477 | 371 | 78% | 1.89 |
| Logistics (Lo) | 133 | 111 | 83% | 1.33 |
| Logistics Coeff (LoCo) | 16 | 16 | 100% | 2.75 |
| Water (Wa) | 111 | 111 | 100% | 1.00 |
| Energy Supply (ES) | 34 | 33 | 97% | 1.79 |
| Energy Sup Coeff (ESCo) | 12 | 12 | 100% | 6.00 |
| Convert Factors (CF) | 2 | 2 | 100% | 18.00 |
| Out: Energy Dem (SSED) | 185 | 12 | 6% | 3.32 |
| Out: Energy Sup (SSES) | 244 | 48 | 20% | 4.40 |
| Out: Logistics (SSLo) | 67 | 0 | 0% | 3.99 |
| Out: Pass Trans (SSPT) | 366 | 0 | 0% | 3.71 |
| Out: Socio-Econ (SSSE) | 14 | 0 | 0% | 4.21 |
| Out: Water (SSW) | 264 | 75 | 28% | 4.08 |
| Project Outputs (Out) | 6 | 0 | 0% | 14.83 |

Figure 4 Per discipline metrics calculated from a calculation graph extracted from a model slice for annual per capita carbon emissions.

From the number of inputs in each model we gain an indication of each discipline's data demands. Finally we see that although each discipline has both an input and an output model, this demarcation is not strictly observed in all disciplines. The inputs within output models are of particular concern; though these are sometimes conversion factors or calculation options. Similarly many input models have up to 40% non-input (i.e.

calculation) cells. This is acceptable since they summarise the input data for use in other models (e.g. total land use).

Together with metrics for the most referenced input data and sub-calculations, these multidisciplinary metrics give a key overview of the model focus as well as aiding the maintenance of the model by checking whether design rules are followed. This is particularly important in engineering models where model structure is constantly evolved by practitioners.

## 8. Discipline Coupling

| | 01LU | 02SE | 03PT | PTCo | 04ED | 05Lo | LoCo | Wa | 08ES | ESCo | CF | SS_Lo | SS_PT | SS_SE | SS_ES | SS_W | SS_ED | Out |
|---|---|---|---|---|---|---|---|---|---|---|---|---|---|---|---|---|---|---|
| 01LU | 19 | 5 | 0 | 0 | 0 | 0 | 0 | 0 | 0 | 0 | 0 | 0 | 0 | 12 | 0 | 15 | 53 | 0 |
| 02SE | 0 | 19 | 0 | 0 | 0 | 0 | 0 | 0 | 0 | 0 | 0 | 0 | 0 | 23 | 0 | 5 | 0 | 0 |
| 03PT | 0 | 0 | 60 | 0 | 0 | 0 | 0 | 0 | 0 | 0 | 0 | 0 | 210 | 0 | 0 | 0 | 0 | 0 |
| PTCo | 0 | 0 | 0 | 81 | 0 | 0 | 0 | 0 | 0 | 0 | 0 | 0 | 180 | 0 | 0 | 0 | 0 | 0 |
| 04ED | 0 | 0 | 0 | 0 | 424 | 0 | 0 | 0 | 0 | 0 | 0 | 0 | 0 | 0 | 0 | 0 | 53 | 0 |
| 05Lo | 0 | 0 | 0 | 0 | 0 | 44 | 0 | 0 | 0 | 0 | 0 | 89 | 0 | 0 | 0 | 0 | 0 | 0 |
| LoCo | 0 | 0 | 0 | 0 | 0 | 0 | 0 | 0 | 0 | 0 | 0 | 44 | 0 | 0 | 0 | 0 | 0 | 0 |
| Wa | 0 | 0 | 0 | 0 | 0 | 0 | 0 | 0 | 0 | 0 | 0 | 0 | 0 | 0 | 0 | 111 | 0 | 0 |
| 08ES | 0 | 0 | 0 | 0 | 0 | 0 | 0 | 0 | 1 | 0 | 0 | 0 | 0 | 59 | 0 | 0 | 0 | 0 |
| ESCo | 0 | 0 | 0 | 0 | 0 | 0 | 0 | 0 | 0 | 0 | 0 | 0 | 72 | 0 | 0 | 0 | 0 | 0 |
| CF | 0 | 0 | 0 | 0 | 0 | 0 | 0 | 0 | 0 | 0 | 1 | 0 | 0 | 0 | 35 | 0 | 0 | 0 |
| SS_Lo | 0 | 0 | 0 | 0 | 0 | 0 | 0 | 0 | 0 | 0 | 0 | 44 | 0 | 0 | 0 | 0 | 45 | 0 |
| SS_PT | 0 | 0 | 0 | 0 | 0 | 0 | 0 | 0 | 0 | 0 | 0 | 0 | 453 | 0 | 0 | 0 | 60 | 3 |
| SS_SE | 0 | 0 | 0 | 0 | 0 | 0 | 0 | 0 | 0 | 0 | 0 | 0 | 0 | 6 | 0 | 7 | 0 | 5 |
| SS_ES | 0 | 0 | 0 | 0 | 0 | 0 | 0 | 0 | 0 | 0 | 0 | 0 | 0 | 424 | 0 | 0 | 1 | 59 |
| SS_W | 0 | 0 | 0 | 0 | 0 | 0 | 0 | 0 | 0 | 0 | 0 | 0 | 0 | 0 | 449 | 0 | 6 | 0 |
| SS_ED | 0 | 0 | 0 | 0 | 0 | 0 | 0 | 0 | 0 | 0 | 0 | 0 | 35 | 0 | 0 | 175 | 0 | 12 |
| Out | 0 | 0 | 0 | 0 | 0 | 0 | 0 | 0 | 0 | 0 | 0 | 0 | 0 | 0 | 0 | 0 | 0 | 5 |

**Figure 5 Discipline coupling matrix shows discipline communication in the IRM model. Matrix should be read ``x values in *row* model are used by *column* model''. Circles indicate the presence of indirect references.**

Since multidisciplinary models contain sub-models from many different disciplines, we consider the interconnections between these disciplines as shown by data dependencies in spreadsheet formulas.

As a concrete example, one hypothesis proposed by the IRM engineers was that the transport model was not connected to the land-use model (since it uses software external to the spreadsheet). In order to test this, a discipline coupling matrix was created (Figure 5). This is calculated by considering all edges in the calculation graph and entering them into the matrix according to which disciplines they are from/to (effectively recording cross worksheet reference in formulas). We see the passenger transport and logistics models (PT, Lo and their Coefficients) are indeed not directly connected to the land-use (LU) model, thus confirming the IRM engineers' hypothesis.

Due to the breakdown of inputs/output models within disciplines we see that the top right quadrant covers output models reading from input models. The bottom left quadrant covers input models reading from output models (which shouldn't and doesn't occur). Much of the model complexity is found in the bottom right quadrant with interconnected calculation models. The diagonal shows sub-model complexity via internal references. We also consider indirect references (reference via another model) and note the primacy of the energy demand and supply models which reference almost all other disciplines.

This is a key technique for considering multidisciplinary engineering models and enables validation that the spreadsheet created matches a conceptual model of communication dataflow.

# 9. Coupling Metrics

Given the number of sub-models comprising the IRM model, there is much similarity between large spreadsheets and large software programs. Considering each discipline's models as separate code packages we can apply standard software engineering code metrics [Martin, 2006] to the discipline coupling matrix. Such metrics are in a similar vein to [Hermans, 2012] but are calculated at the worksheet level by treating worksheets as packages (inspired by their relationship to disciplines).

|  | Afferent Coupling (Responsibility) | Efferent Coupling (Independence) | Instability |
|---|---|---|---|
| Land Use | 4 | 0 | 0% |
| Socio Economic | 2 | 1 | 33% |
| Passenger Trans | 1 | 0 | 0% |
| Pass Trans Coeff | 1 | 0 | 0% |
| Energy Demands | 1 | 0 | 0% |
| Logistics | 1 | 0 | 0% |
| Logistics Coeff | 1 | 0 | 0% |
| Water | 1 | 0 | 0% |
| Energy Supply | 1 | 0 | 0% |
| Energy Sup Coeff | 1 | 0 | 0% |
| Convert Factors | 2 | 0 | 0% |
| Out: Energy Dem | 2 | 6 | 75% |
| Out: Energy Sup | 2 | 3 | 60% |
| Out: Logistics | 1 | 3 | 75% |
| Out: Pass Trans | 2 | 2 | 50% |
| Out: Socio-Econ | 2 | 2 | 50% |
| Out: Water | 1 | 5 | 83% |
| Project Outputs | 0 | 4 | 100% |

Figure 6 Software engineering metrics normally applied to large software projects [Martin06] are applied to multidisciplinary models to gain insight into model maintainability and stability to change.

We calculate measures of a model's *responsibility to* and *independence from* other models in terms of the data they provide and consume from other models. We use these to compute *instability* to *model change* to identify which models are most likely to cause difficulty for project adaptation.

Firstly, we compute a model's *afferent* coupling [Martin, 2006] by counting the number of discipline models (worksheets) which reference cells in the given model (worksheet). This gives a measure of the responsibility of a model to other models. Models with high *afferent* coupling are less easy to adapt to new projects as changes must avoid breaking its dependant's expectations.

Secondly, we compute *efferent* coupling [Martin, 2006] by counting the number of models (worksheets) which cells in a given model (worksheet) reference. This gives a measure of the independence of the model, with lower scores considered more independent. Models with poor independence are likely to be affected by changes in other models.

Finally we compute a measure of a model's instability to change [Martin, 2006] as follows, where 0% is stable and 100% is unstable.

$$\frac{\text{efferent}}{\text{afferent} + \text{efferent}} = \text{Instability}$$

Figure 6 shows the results for the IRM model. As expected most discipline input models are highly independent and not likely to be affected by changes to other models. Conversely the output models have varying levels of dependence on other models and so have higher levels of instability. This indicates they are more likely to be affected by model changes, particularly as the model evolves. Instability also coarsely identifies flows of effects from changes in input.

These metrics allow engineers to quantify the difficulty and risks involved in making changes to a given model and how likely these changes are to affect other disciplines' models; frequent reference to these metrics should create more modular model which are less costly to adapt, in part by identifying erroneous or poorly planned connections between worksheets.

## 10. Sensitivity Analysis

One common engineering task is to optimise a design for a given KPI, for example annual per capita carbon emissions. This is difficult since the designer must identify all input cells which affect the KPI, consider their ranges and then attempt to find combinations of values which optimise the KPI whilst considering the impacts of doing so.

In support of this we created tools to apply a sensitivity analysis to a slice of the spreadsheet corresponding to all of the cells involved in calculating a KPI of interest. A sensitivity analysis identifies the input factors to which the KPI is most sensitive to changes in. This enables the designer to focus upon the subset of inputs which have the most effect on the KPI, increasing their productivity.

| Variable | Normalised Sensitivity for $CO_2e$ Emissions Per Capita | | |
|---|---|---|---|
| | Total | Non-Domestic Buildings | External Transport |
| FuelType Petrol City Car | 100 | 0 | 100 |
| CO2 emissions from gas combustion | 91 | 100 | 0 |
| FuelType Electric Heavy Rail | 78 | 0 | 78 |
| District Heat Demand - Gas Boiler | 71 | 73 | 0 |
| District Heat Efficiency - Gas Boiler | 71 | 73 | 0 |
| Gas Network Efficiency | 68 | 83 | 0 |
| Gas Network Demand | 62 | 78 | 0 |
| Electricity Demand from CHP | 57 | 68 | 0 |
| CH4 emissions from biomass | 47 | 52 | 0 |
| Efficiency of Heat from biomass | 46 | 55 | 0 |

Figure 7 A sensitivity analysis identifies the variable with most scope to impact a KPI. Results normalised to the most impactful variable. We show impact upon total percapita carbon and side effects on some constituent parts of this figure.

In contrast to many tools we use Design of Experiments techniques to create efficient experiments for interrogating the sensitivity of a model. These techniques take a set of factors (inputs), which affect the output of interest, along with the maximum and minimum value each factor can take (set by the practitioner). A series of model runs is then constructed with varying combinations of factors set at their maximum or minimum levels. These are then run and the results analysed. We use a Plackett-Burman (PB) sensitivity analysis [Plackett, 1946] due to its computational efficiency, which comes at the cost of insight only into the effects of factors and not their interactions.

As an example we consider the annual per capita carbon emissions KPI, extract the corresponding model slice and identify its numeric inputs. For each input the maximum and minimum range of the variable is established with engineers from the appropriate discipline. Note that not all numeric inputs are variable e.g. conversion factors. This produced 933 parameters for a sensitivity analysis; Figure 7. shows the results. This requires 2,563 simulation runs, the Excel-Sensitivity tool runs one run per 0.72 seconds on a Quad Core (Intel i7 720QM) machine, running four experiments concurrently.

Since we can test the sensitivity of more than one KPI to the same set of factors at very little extra cost, we explore side effects on the breakdown of the total per capita $CO_2e$ emissions. This gives insight into the relative importance of each sub-metric to the total and what scope there is for affecting each. For example, different fuel type metrics affect the total $CO_2e$ emissions and the transport KPIs but do not affect the non-domestic buildings sub-metric. Interestingly, we see that district heating has a surprisingly high effect on the carbon efficiency, as do Combined Heat and Power (CHP) systems should they be included in the masterplan. These results can be broken down by discipline for more detailed insight.

In conclusion we identify these benefits of a sensitivity analysis on an engineering model:

- **Design Insight** - The designer gains knowledge of the design space, the interactions between the design parameters and the output KPIs of interest allowing a focusing of effort upon only those variables the KPIs is most sensitive to and similarly gaining insight into side effects of changes on other KPIs.
- **Design Space Exploration** - Whilst running the analysis we automatically create and evaluate several thousand designs. Exploration of these allows designers to quickly understand potential configurations and directions for design improvements.
- **Identification of effects of assumptions** - Within most engineering models there are a large number of calculation assumptions. For example, the carbon emissions of buses per passenger kilometre. If included within a sensitivity analysis (the max/min values determining the confidence interval of the assumption) the engineer gains understanding of the relative importance of the assumptions and the respective effects of error margins; enabling focus on refining model uncertainty which will have most impact. This also mitigates risks in analysis accuracy since unexpectedly sensitive inputs are identified.

## 11.    IRM Evolution

One interesting use of EAM is its repeated application to a model, particularly as it is adapted to meet the requirements of new projects. We explored the application of the EAM toolkit with three IRM models developed over a number of years from a concept case study to a globally used tool; demonstrating the transferability and scalability of the methodology and tools.

Figure 8 shows such application of EAM to Arup's IRM model. The size of the model has increased dramatically as more detail and accuracy have been added to the model. This is partly due to the most recent IRM model containing data tables localised to geographical regions. The increase in size also reflects an increase in complexity, as noted by the number of Excel functions called within the model. All figures in the table refer to the slice of the model corresponding to annual per capita carbon emissions. The complexity increase compounds the problems discussed in Section 2, highlighting the need for computational support.

| Formulas Used in IRM models | | | | | |
|---|---|---|---|---|---|
| IRM 2008 | | IRM 2009 | | IRM 2011 | |
| 1,234 Cells | | 2,357 Cells | | 37,926 Cells | |
| 2,360 References | | 3,404 References | | 253,222 References | |
| SUM | 79 | SUM | 176 | IF | 5250 |
| | | IF | 99 | MATCH | 2714 |
| | | TYPE | 81 | HLOOKUP | 2714 |
| | | | | ROUNDUP | 1717 |
| | | | | ISERROR | 1357 |
| | | | | SUM | 1223 |
| | | | | VLOOKUP | 198 |
| | | | | SUMIF | 78 |
| | | | | AND | 57 |
| | | | | ISNUMBER | 28 |
| | | | | Misc | 7 |

**Figure 8 We applied the EAM [Liang11] toolkit to three different IRM models ranging from a concept model to a fully developed model to a globally used geographically localized tool.**

From computing the metrics discussed in Sections 7-9 for each model, we identified the change in focus over time from water modelling through to energy and carbon models by considering the change in complexity and connectivity between the disciplines within the model. These insights demonstrate transferability of the approach. We have been able to apply the process and tools to an Arup vertical transportation model from start to end within one working day reporting valuable insight into the model which was accepted by its expert maintainer.

## 12.     Further Work

Firstly, we see that high data demands are a barrier to IRM adoption. However, it may be possible for automatic or semi-automatic methods to be applied to the calculation graph to attempt to produce an abstracted version of the model with fewer data requirements. Methods such as sensitivity analysis could be used to identify parts of the model for removal which have limited impact upon the accuracy of the overall model results.

Secondly, whilst a PB sensitivity analysis gives good insight into the model, other forms of sensitivity analysis might be applied (dependent upon their tractability). One of the more interesting methods would be to apply automatic differentiation tools to the overall model formula allowing accurate insight into the multi-variate sensitivities of the model.

Finally, given the formalisation of discipline specific implicit knowledge behind the variable ranges for a sensitivity analysis, it would be interesting to use these as the constraints to an optimisation engine performing constraint based optimisation upon the model. Of course such optimisation would never replace an engineer's insight into which combinations of variable values are practicable but could serve as a valuable decision support tool within IRM models and other engineering models. Particularly the authors see this work and its methodology as being applicable to spreadsheet systems in other domains (e.g. financial) and particularly groups of interlinked spreadsheets.

## 13.     Conclusions

The case study presented demonstrates the need and the value of computational tools in understanding complex multidisciplinary models. The techniques explored aid practitioners in model comprehension, optimisation and evolution; as evidenced by exploring Arup's IRM model as a representative model and the aid given to practitioners. Many of whom have no formal programming experience with model development tasks. Model slicing allows reduction of model complexity to show only the salient points. Interactive exploration of the model as a calculation graph valuably enables users to build a mental model of how the calculation works. Model metrics are an interesting and valuable way of gaining detailed insight into the model and its composition. Metrics pertaining to the multidisciplinary nature of the model give higher level insight into interdisciplinary communication. Finally, sensitivity analysis is a valuable technique for understanding the relative importance of hundreds of input variables when seeking to optimise for a given KPI or checking model assumptions. Repeated application of EAM to a model clearly identifies changes in model composition and focus. EAM and its techniques can be applied more widely than IRM models and have been applied to other confidential engineering models. Indeed the approach has been applied to a system of some 500 connected spreadsheet files.

In conclusion:
- We present a methodology and tool suite for systematic, automatic analysis of large spreadsheet-based models with novel metrics to assess internal communication and integrated sensitivity analysis to aid practitioners in optimisation.

- We applied this methodology with a focus upon multidisciplinary engineering assessment models, a model type not widely studied within literature.

- We demonstrated the methodology's application through practical case studies with an industrial multidisciplinary sustainability model, identifying insight for practitioners and model evolution over three generations.

The authors would like to gratefully acknowledge the support of the EPSRC and Arup in funding and supporting this work.

# 14. References

Ayaz, E. and Levitas, J. (2008), "Spatially Linked Integrated Resource Management (IRM): A Tool to Inform Eco-city Planning", Proceedings of the 8th International Eco-city Conference (Eco-city 2012).

Blayney, P. J. (2006), "An Investigation of the Incidence and Effect of Spreadsheet Errors Caused by the Hard Coding of Input Data Values into Formulas".

Clermont, M. (2005), "Heuristics for the Automatic Identification of Irregularities in Spreadsheets", SIGSOFT Software. Eng. Notes.

Kankuzi, B. and Ayalew, Y. (2008), "An End-user Oriented Graph-based Visualization for Spreadsheets", Proceedings of the 4th International Workshop on End-user Software Engineering.

Hermans, F., Pinzger, M., and Deursen, A. (2011), "Supporting Professional Spreadsheet Users by Generating Leveled Dataflow Diagrams", 33rd International Conference on Software Engineering (ICSE), pages 451–460.

Hermans, F., Pinzger, M. and Deursen A. (2012), "Detecting and Visualizing Inter-worksheet Smells in Spreadsheets", 34th International Conference on Software Engineering (ICSE), pages 441–451.

Hodnigg, K. and Mittermeir, R. T. (2008), "Metrics-Based Spreadsheet Visualization Support for Focused Maintenance", Proceedings of EuSpRiG 2008, pages 79–94.

Kepran, H. (2002), "Creating a Framework for Integrated Resource Management" ICLEI 2002.

Liang, H. and Birch, D. (2011), "Extraction and Analysis Methodology for Supporting Complex Sustainable Design", Proceedings of the18th International Conference on Engineering Design (ICED11).

Martin, R. and Martin, M. (2006), "Agile Principles, Patterns, and Practices in C#", Prentice Hall.

Page, J., Grange, N. and Kirkpatrick, N. (2008), "The Integrated Resource Management (IRM) Model Guidance Tool for Sustainable Urban Design", Proceedings of the 25th Conference on Passive and Low Energy Architecture (PLEA08).

Panko, R. R. (2008), (Revised edition) (2008), "What We Know About Spreadsheet Errors", Journal of End User Computing, pages 15–21.

Panko, R. R. and Aurigemma, S. (2010), "Revising the Panko-Halverson Taxonomy of Spreadsheet Errors Decision". Support Sys, pages 235-244.

Plackett, R.L and Burman, J.P. (1946), "The Design of Optimum Multifactorial Experiments", Biometrika, pages 305-325.

Reichwein, J., Rothermel, G. and Burnett, M. M. (1999), "Slicing Spreadsheets: An Integrated Methodology for Spreadsheet Testing and Debugging", Proceedings of the 2nd Conference on Domain-specific Languages, pages 25-38.

Shiozawa, H., Okada, K. and Matsushita, Y. (1999), "3D Interactive Visualization for Inter-Cell Dependencies of Spreadsheets", IEEE Symposium on Information Visualization, pages 79-82.

Weiser, M. (1981), "Program Slicing", Proceedings of the 5th International Conference on Software Engineering. pages 439–449.